\documentclass[]{aastex63}
\usepackage{float}
\shorttitle{FFA search in GHRSS data}
\shortauthors{ }
\graphicspath{{./}{figures/}}

\begin{document}

\title{The GMRT High Resolution Southern Sky Survey for pulsars and transients - III: searching for long period pulsars}

\author{S.~Singh}
\affiliation{National Centre for Radio Astrophysics, Tata
Institute of Fundamental Research, Pune 411 007, India}

\author{J.~Roy}
\affiliation{National Centre for Radio Astrophysics, Tata
Institute of Fundamental Research, Pune 411 007, India}

\author{U.~Panda}
\affiliation{National Centre for Radio Astrophysics, Tata
Institute of Fundamental Research, Pune 411 007, India}

\author{B.~Bhattacharyya}
\affiliation{National Centre for Radio Astrophysics, Tata
Institute of Fundamental Research, Pune 411 007, India}

\author{V.~Morello}
\affiliation{Jodrell Bank Centre for Astrophysics,
School of Physics and Astronomy, The University of Manchester,
Manchester M13 9PL, UK}

\author{B.~W.~Stappers}
\affiliation{Jodrell Bank Centre for Astrophysics,
School of Physics and Astronomy, The University of Manchester,
Manchester M13 9PL, UK}

\author{P.~S.~Ray}
\affil{U.S. Naval Research Laboratory, Washington, DC 20375, USA}
\author{M.~A.~McLaughlin}
\affil{Department of Physics and Astronomy, West Virginia University, Morgantown, WV 26506, USA}
\affil{Center for Gravitational Waves and Cosmology, West Virginia University, Chestnut Ridge Research Building, Morgantown, WV 26505, USA}






\begin{abstract}

Searching for periodic non-accelerated signals in the presence of ideal white noise using the fully phase-coherent Fast Folding Algorithm (FFA) is theoretically established as a more sensitive search method than the Fast Fourier Transform (FFT) search with incoherent harmonic summing. In this paper, we present a comparison of the performance of an FFA search implementation using \texttt{RIPTIDE} and an FFT search implementation using \texttt{PRESTO}, over a range of signal parameters with white noise and with real telescope noise from the GHRSS survey with the uGMRT. We find that FFA search with appropriate de-reddening of time series, performs better than FFT search with spectral whitening for  
long period pulsars in real GHRSS noise conditions. We describe an FFA search pipeline implemented for the GHRSS survey looking for pulsars over a period range of 0.1 s to 100 s and up to a dispersion measure of 500 pc cm$^{-3}$. We processed GHRSS survey data covering $\sim$ 1500 degree$^2$ of the sky with this pipeline. We re-detected 43 known pulsars with better signal-to-noise in the FFA search than in the FFT search. We also report the discovery of two new pulsars, including a long period pulsar having a short duty-cycle with this FFA search pipeline. The population of long period pulsars with periods of several seconds or higher can help to constrain the pulsar death-line.

\end{abstract}



\section{Introduction} \label{sec:intro}

Understanding the mechanism of radio emission from neutron stars is still an unresolved problem (\citet{Beskin_2018}, \citet{Mitra_2017} and \citet{Cerutti_2016}). However, significant progress has been made over the last decades, aided by discoveries of new pulsars exhibiting interesting phenomena like drifting \citep{drifting_drake}, nulling \citep{Backer_1970} and mode changing \citep{bartel_1982} carrying imprints of the underlying emission processes.
For a given magnetic field configuration, standard emission models predict a critical value of magnetic field strength and hence a period derivative corresponding to the spin period of a pulsar below which radio emission ceases (\citet{chen_and_rutherman}, \citet{RS75}). These values of pulse period (P) and period derivative ($\dot{P}$) trace a curve on the $P-\dot{P}$ diagram, called the death-line \citep{chen_and_rutherman}. This line varies significantly for different magnetic field configurations and emission models, and spans a region on the $P-\dot{P}$ diagram, called the death-valley (\citet{single_park} and \citet{chen_and_rutherman}).
The discovery of one of the long period pulsars, J2144$-$3933 \citep{young_1999} with a spin period of 8.5 s, and its location on the $P-\dot{P}$ diagram challenges all the existing pulsar radio emission models (\citet{zhang2000}, \citet{chen_and_rutherman}). 
Although the discoveries of a 23.5 s pulsar J0250$+$5854 \citep{Tan_2018} and a 12.1 s pulsar J2251$-$3711 \citep{Morello_2020} continue to constrain the emission models, J2144$-$3933, the 8.5 s pulsar still provides the most stringent constraints due to its smaller period derivative.
This illustrates the importance of finding such long period pulsars located close to the death-line.

Over the last decade, the number of millisecond pulsars (MSPs) has increased by four fold\footnote{http://astro.phys.wvu.edu/GalacticMSPs/GalacticMSPs.txt}, whereas there has been only a marginal increase ($\sim$50\% \footnote{https://www.atnf.csiro.au/research/pulsar/psrcat/}) in the number of long period pulsars (P $>1$ s). 
While the highly efficient search of millisecond pulsations
following the Fermi gamma-ray unassociated sources\footnote{https://confluence.slac.stanford.edu/display/GLAMCOG/Public+List+of+LAT-Detected+Gamma-Ray+Pulsars} could significantly aid the MSP population, the lack of long period pulsars can be caused by both intrinsic and observational biases.  
Due to the $P^{-0.5}$ dependence of the radius of the opening angle of the polar cap \citep{skrzypczak}, narrow radio beam of long period pulsars could miss the observer line-of-sight \citep{young_1999}, introducing an intrinsic bias against the long period pulsars. 
One of the observational biases against long period pulsars comes from the fact that the majority of the pulsar surveys spend a few minutes of integration time per pointing, which only contain a few pulses from long period pulsars \citep{Parent_2018}. There are also biases introduced by search algorithms. \citet{Kondratiev_2009} and \citet{Cameron_2018} reported that the conventional FFT (Fast Fourier Transform) based search method (e.g. \texttt{PRESTO}, \cite{Ransom_2002}) is less sensitive for long period pulsars, even in ideal white noise conditions. This demands the exploration of different periodicity search methods and an increase in the integration time per pointing, in order to mitigate the possible observational biases against long period pulsars.

As shown in previous studies (e.g. \citet{Parent_2018}, \citet{Cameron_2018} and \citet{Heerden_2016}), searching for periodic signals in the frequency domain can be affected by red noise. The slowly varying instrumental gain can contribute increased power in the lower part of the power spectrum (e.g. up to a few tens of Hz for the GMRT time-domain data). 
This can limit the sensitivity of the search for longer period pulsars.
Since the signal power in a FFT-based periodicity search is accumulated over a limited harmonic space, incoherently summing up power in the first few harmonics (as conventionally done for FFT based searches), leads to a loss in sensitivity.   
In another method for periodicity search, called the Fast Folding Algorithm (FFA, \citet{staelin}), the periodic signal is searched by folding the time series at multiple closely spaced trial periods. Folding in the time domain is a coherent process and preserves all of the signal power. The latest study on the sensitivity of FFA and FFT search \citep{RIPTIDE} shows that in ideal white noise, FFA search is much more sensitive than FFT search for small duty-cycle pulsars, irrespective of the period. However these theoretical predictions applicable to ideal white noise conditions need to be tested in real noise conditions (affected by red noise and radio frequency interference). 

The computational speed of FFA search depends on the underlying period being searched and gets more expensive while searching for pulsars with smaller periods. The availability of increased computing power in the past few years has enabled many major pulsar surveys to apply FFA search algorithms (e.g. SUPERB survey\footnote{https://sites.google.com/site/publicsuperb/discoveries}, PALFA survey\footnote{http://www2.naic.edu/~palfa/newpulsars/index.html} and HTRU-S survey\footnote{https://sites.google.com/site/htrusouthdeep/} ). Some of the available FFA search packages are \texttt{ffancy} \citep{Cameron_2018}, \texttt{ffaGo} \citep{Parent_2018} and \texttt{RIPTIDE} \citep{RIPTIDE}. These packages are developed to efficiently find pulsars at the higher end of the periodicity range while mitigating the red noise. However, for the current implementation of pulsar search with the GMRT, we have opted for the \texttt{RIPTIDE} package. \texttt{RIPTIDE} uses a faster variant of FFA along with a better candidate selection method \citep{RIPTIDE}. We are searching for pulsars and fast transients with the GMRT High Resolution Southern Sky (GHRSS\footnote{http://www.ncra.tifr.res.in/$\sim$bhaswati/GHRSS.html}, \citet{GHRSS2}, \citet{GHRSS1}) survey, away from the Galactic plane. Using an FFT based search, the GHRSS survey has already discovered 21 pulsars. In this paper, we describe the FFA-based search methodology for the GHRSS survey along with a few initial discoveries. In Section \ref{sec:FFA vs FFT}, we perform a comparative study of FFA and FFT search over a range of different parameters related to noise conditions (simulated white noise, real GHRSS noise, and simulated red noise conditions) and the pulsar signal (period of pulsed signal, width of pulses, and shape of pulses). The GHRSS survey and its prospects of finding long period pulsars using FFA search on the archival data as well as in new data from ongoing observations, the data processing pipeline, and optimisations of its parameters are discussed in Section \ref{sec:processing}.  Results, including new discoveries, are presented in Section \ref{sec:discoveries}. Finally, in Section \ref{sec:discussion} we summarize the outcomes of this work.

\section{Comparison of FFA search and FFT search} \label{sec:FFA vs FFT}
\citet{RIPTIDE} derived the mathematical framework for a comprehensive search methodology, which is to phase coherently fold the time series and then correlate it with a pulse profile template. In a blind search, in absence of information about the period and the shape of pulse, we need to fold at a number of closely-spaced trial periods and use a suitable pulse-shape for profile template. In general, boxcars of different sizes are used as profile templates to minimize the computational cost.  \citet{RIPTIDE} have done a detailed and rigorous theoretical analysis of FFA and FFT search sensitivity, where the behavior of FFA and FFT search sensitivity as a function of period and duty-cycle has been predicted. These predictions are expected to hold in ideal white noise conditions but tests are needed to assess the performance of FFA and FFT search in real telescope data (real noise) containing non-gaussian noise contributed by radio frequency interference (RFI) and red noise.  

Comparisons of FFA and FFT search sensitivity in ideal white noise and real noise conditions have been done by \citet{Parent_2018}, \citet{Cameron_2018}, and \citet{Kondratiev_2009}. All these studies concluded that even in ideal white noise conditions, FFA outperforms FFT search for long periods. \citet{Parent_2018} found that with real noise data from the PALFA survey, FFA is more sensitive than FFT search for any dispersion measure (DM, accounts for free electron density integrated along the line of sight) and duty cycle if period is large enough ($\gtrsim$8~s). We carried out a similar test in white noise and GHRSS noise conditions to determine the behavior of FFA and FFT search for our survey while testing the predictions made by \citet{RIPTIDE} about the two search methods. 

For our test, we selected five GHRSS noise files over 300$-$500 MHz with 4096 channels at 81.92 $\mu$s time resolution (refer to Section \ref{sec:processing} for survey specifications).
These five noise files have different spectro-temporal features generated from RFIs as well as different red noise contributions but do not contain any known pulsar signal.
We also simulated five white noise files using \texttt{fast\_fake} of \texttt{SIGPROC\footnote{https://github.com/SixByNine/sigproc}} (having similar time-frequency resolution) of 10 mins duration, which is the pointing duration of the GHRSS survey. 
To compare the FFA and FFT search over various signal parameters, we injected pulses (using \texttt{inject\_pulsar} of \texttt{SIGPROC}) at a range of periods, duty-cycles, and pulse shapes in these 10 noise files (5 real GHRSS noise and 5 white noise). Then we performed FFA search and FFT search on these simulated pulsar data files using \texttt{RIPTIDE} \citep[version: 0.0.1,][]{RIPTIDE} and \texttt{PRESTO} \citep[version: 07Dec16,][]{Ransom_2002} respectively. We de-dispersed the filterbank files using \texttt{prepsubband} of \texttt{PRESTO}. Then, FFA and FFT searches were performed on the dedispersed time-series. We note that even with the same mean (phase averaged) signal-to-noise (S/N) used for injection in \texttt{inject\_pulsar} (decided by scaling factor \texttt{s $=$ 1.5}) for all the real noise simulations, we obtained different S/Ns in the folded profile for different real noise inputs contaminated by baseline variations and RFIs. Thus, we normalise (refer to respective sections for details of normalisation) the detection S/Ns for each of the noise cases separately to investigate the variations of detection S/Ns with pulsar parameters. 

For FFA search, we use \texttt{ffa\_search} of \texttt{RIPTIDE} on the time-series with \texttt{bins\_min=960} and \texttt{bins\_max=1040}, which gives a range for number of bins used in the folded profile. This decides the duty-cycle resolution (the minimum duty-cycle which can be searched for) and down-sampling factor (refer to \citealt{RIPTIDE}). For example, the above values of \texttt{bins\_min} and \texttt{bins\_max} keep the duty-cycle resolution close to $0.1\%$.
Before the FFA search is performed, a running median filter is used to remove the baseline variations. The window size of the running median filter is kept at more than twice the expected on-pulse width of the pulsar in order to avoid attenuating the pulse. We use different feasible values for \texttt{rmed\_width} (given in relevant sections for each simulation), which decides the window size of the running median filter.    
The \texttt{ffa\_search} routine generates a periodogram, which is S/N as a function of trial periods and widths (width of boxcars). Periodic signals manifest as local peaks in the periodogram, and a peak detection algorithm called \texttt{find\_peaks} is used to identify these peaks above 7$\sigma$ threshold.

In FFT search, we took the Fourier transform of dedispersed time series using \texttt{realfft} of \texttt{PRESTO} \citep{Ransom_2002}. Then we used \texttt{rednoise} module of \texttt{PRESTO} on the power spectra with default choices of parameters (\texttt{startwidth}=6, \texttt{endwidth}=100, \texttt{endfreq}=6 Hz) for spectral whitening. The parameters \texttt{startwidth} and \texttt{endwidth} define the range of block sizes for median estimation in terms of Fourier frequency bins for dereddening (refer to Fig. 4 of \cite{Heerden_2016}). The above values translate to $\sim$0.01 Hz for \texttt{startwidth} and $\sim$0.17 Hz for \texttt{endwidth} for the GHRSS survey. 
We performed periodicity search using \texttt{accelsearch} of \texttt{PRESTO} with \texttt{zmax=0} (i.e. no acceleration correction), \texttt{numharm=8} and \texttt{flo=0.019} (which defines lowest frequency for the 8th harmonic, i.e. to search up to period $\sim$ 421 s) on the dereddened power spectra. FFT search in the GHRSS survey is also simultaneously looking for mildly accelerated millisecond pulsars \citep{GHRSS1}. Since the maximum frequency bin drift that can be corrected scales linearly with harmonic frequency (Eq. 6.17 in \citet{handbook}), the search acceleration range reduces with increasing harmonic number. Thus, the current FFT processing of the GHRSS survey is limited to summation of 8 harmonics.
These tests are aiming to compare the FFA and FFT search implementations in the GHRSS survey processing, thus we are performing summation of 8 harmonics for these tests. We note that the most significant power spectra peak reported by FFT search is not always at the fundamental frequency and more often for longer period pulsars it reports integer multiple of the fundamental frequency (i.e. harmonic) as the pulsar candidate. 

\subsection{FFA and FFT search performance as a function of pulsar period}\label{sec:period_snr}

\cite{Heerden_2016} reported that in the presence of red noise in real telescope data, the sensitivity of FFT search with spectral whitening gets reduced for long period pulsars. However, FFA-search aided with running median subtraction \citep{RIPTIDE} is expected to provide better sensitivity for longer pulse periods \citep{Kondratiev_2009,Parent_2018}.   
In this section, we aim to assess the sensitivity of FFA (using \texttt{RIPTIDE}) and FFT (using \texttt{PRESTO}) search over a range of pulse periods for white noise and for real telescope noise that is affected by RFI and red noise.

\begin{figure}[p]
    \centering
   \includegraphics[height=6.8 cm, width=18.5 cm]{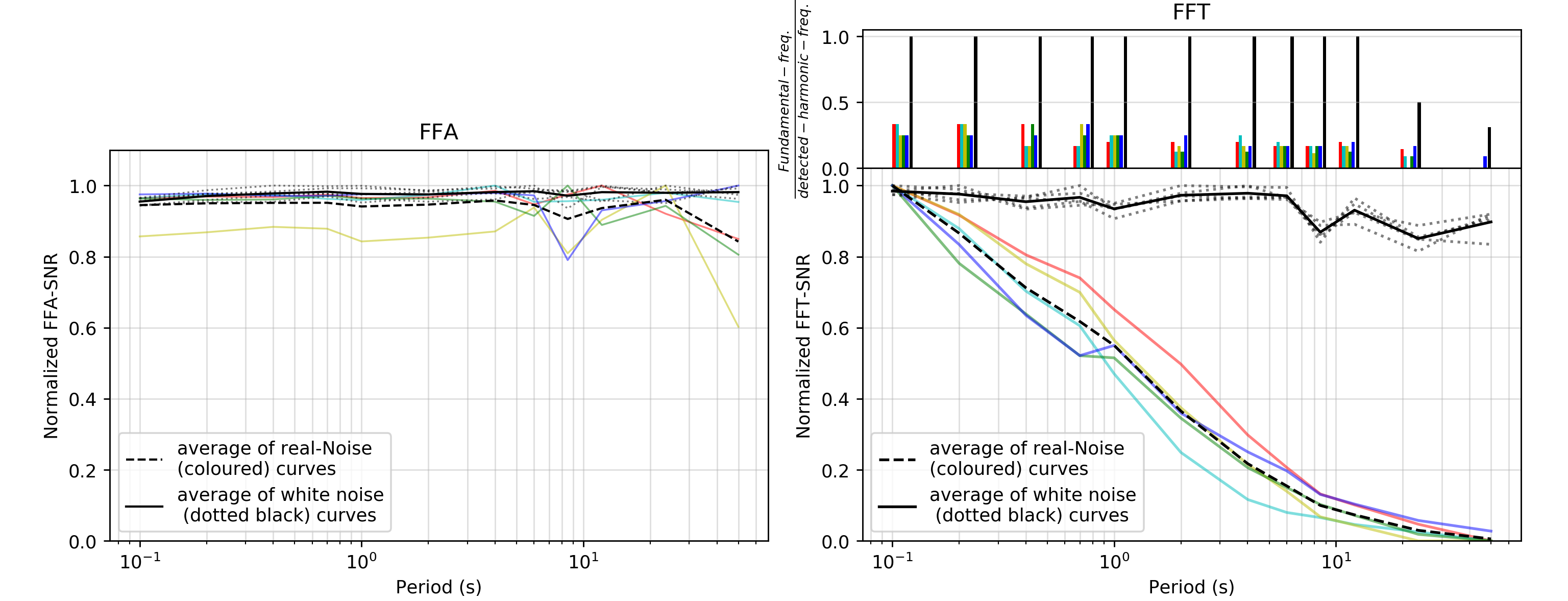}
    \caption{Variations of detection S/N over pulse period for 1\% duty-cycle. Each curve, representing a different noise case, is individually normalised to unit maximum. The faint coloured curves are results in real GHRSS noise, whereas dotted black curves are the results in ideal white noise. The dashed black curve is the average of all real noise cases, and the solid black curve is the average of all white noise curves. For FFA, the detection S/N is almost uniform over the pulse periods, both in white noise as well as in real noise. In the case of FFT search, the detection S/N falls rapidly with period in case of real GHRSS noise conditions, whereas in the case of white-noise the detection S/N remains almost constant with period. In the upper panel of FFT plot, we have the ratio of fundamental frequency and detected harmonic frequency (inverse of detected harmonic) plotted against period for each noise case (having same colour for a given real noise case and black for the average of white noise cases), where we notice a trend of detecting signals at higher harmonic frequencies instead of fundamental for longer periods in real noise cases.}
    \label{fig:snr_period}
\end{figure}

We used 5 real noise files from the GHRSS survey and five simulated white noise files as discussed above, for the comparison of FFA and FFT-search over a range of periods. We generated simulated pulsar data sets of 10 mins duration for each of the 10 noise cases by injecting Gaussian pulses at a range of periods from 0.1 s to 50 s, in 12 period steps and duty-cycle $1\%$ (defined by the full-width-half-maxima, FWHM, of the injected pulse). We sampled the period range 0.1 s $-$ 1 s by 5 trial periods, 1 s $-$ 10 s by 4 trial periods, including one known pulsar period, and then used 3 trial periods in the range 10 s $-$ 50 s, including two known pulsar periods. We performed FFA and FFT search on these data sets and identified the brightest candidates which were matching with the injected pulse periods or their harmonics. We used the detection S/N of these detected candidates to generate curves of detection-S/N versus pulse period.  These curves (for each noise realization) are individually normalised to unit maximum so that we can compare trends of detection-S/N with period in all noise realizations. The value of running median window \texttt{rmed\_width} was kept at 10 s for all the \texttt{RIPTIDE} runs in order to keep the window size sufficiently larger than the on-pulse window considering Gaussian pulse. 
In the \texttt{PRESTO}, de-reddening function \texttt{rednoise} was applied on the power spectra with default choices of parameters. The mean of the real noise curves is expected to represent the results in the average noise condition in the GHRSS data. Fig. \ref{fig:snr_period} shows the plots of S/N versus period in the real noise cases, the average of real noise cases, the white noise cases, and the average of white noise cases. The left panel shows the results of FFA search while the right panel has the results of FFT search.

In the case of FFA, for the white noise cases as well as for the independent real noise cases, normalised detection-S/N remains fairly constant over the entire period range. In the case of FFT, the average of FFT-detection S/N in white noise conditions remains fairly constant over the whole period range. 
While in real noise situations, FFT-S/N falls rapidly with increasing period. For the average GHRSS noise conditions, we observe a sharp decline in FFT-S/N (more than 80\% degradation in S/N as the pulsar period increases to 10 s or higher). In the upper panel of the right plot, we use bar plot to show the variation in the ratio of fundamental frequency and detected harmonic frequency with periods. This ratio of fundamental frequency and frequency of detected harmonic is equivalent to the inverse of harmonic at which signal was detected. S/N versus period curve and inverse of harmonic bar plots are having same colour scheme for a given real noise case, while the black bar plot is for the average of white noise cases. In the real noise cases, we have a trend of detecting signals at higher harmonic frequency for longer periods. As we increase the period of signal, the fundamental and subsequent harmonics shift towards the lower values of modulation frequencies where red noise is dominant and hence we are missing initial harmonics and power in them. While in general, we detect signal at fundamental frequency in white noise. Hence this comparative study in presence of real noise suggests the suitability of FFA-search for longer pulse period. 

\begin{figure}[p]
    \centering
    \includegraphics[height=6.8 cm, width=18.5 cm]{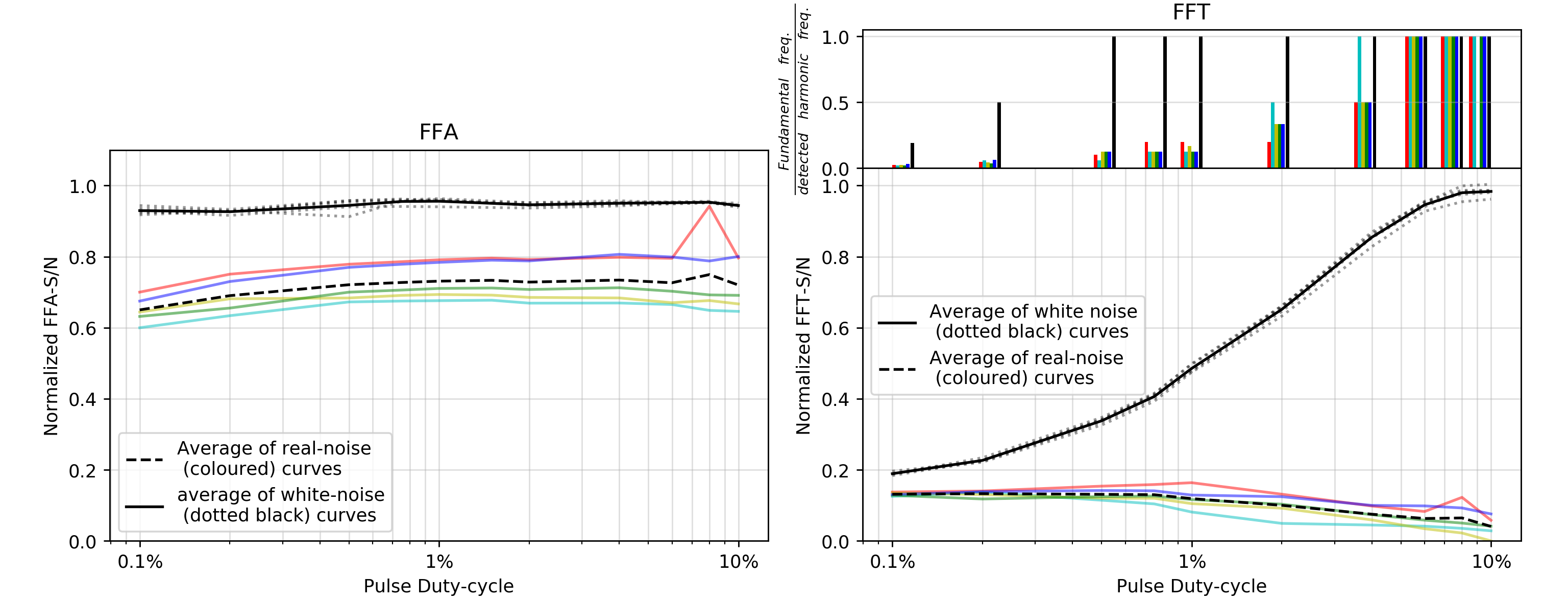}
    \caption{Variations of detection S/N (normalised by S/N achieved from folded profile at nominal period) with pulse duty-cycle for pulse period of 2 s. The faint coloured curves are generated using real GHRSS noise cases, whereas dotted black curves are results in white-noise cases. The dashed black curve is the average of all real noise cases, and the solid black curve is the average of all white noise curves. For FFA, the detection S/N is independent of pulse duty-cycle in both real noise and white noise conditions. FFA search recovers $\sim$ 93\% S/N in case of white noise, while it recovers only $\sim$ 70\% in the real noise cases. Similarly, FFT search also recovers $\sim$100\% of S/N in white noise cases at large duty-cycles ($>$ 8\%), but FFT-S/N remains below 20\% of folded profile S/N in real noise cases for all duty-cycle values. The increase of S/N with duty-cycle seen in presence of white-noise is expected as the fraction of retrieved pulse energy from power spectra (with limited harmonic summation) increases. However, the presence of red noise in real noise simulation limits this signal recovery from power spectra. In the upper panel of the FFT plot, we have the ratio of fundamental frequency and detected harmonic frequency plotted against duty-cycle for each noise case, where we notice a trend of detecting signals at lower harmonics as duty-cycle increases in real noise cases.}
    \label{fig:snr_duty}
\end{figure}

\begin{figure}[p]
    \centering
    \includegraphics[height=6.8 cm, width=18.5 cm]{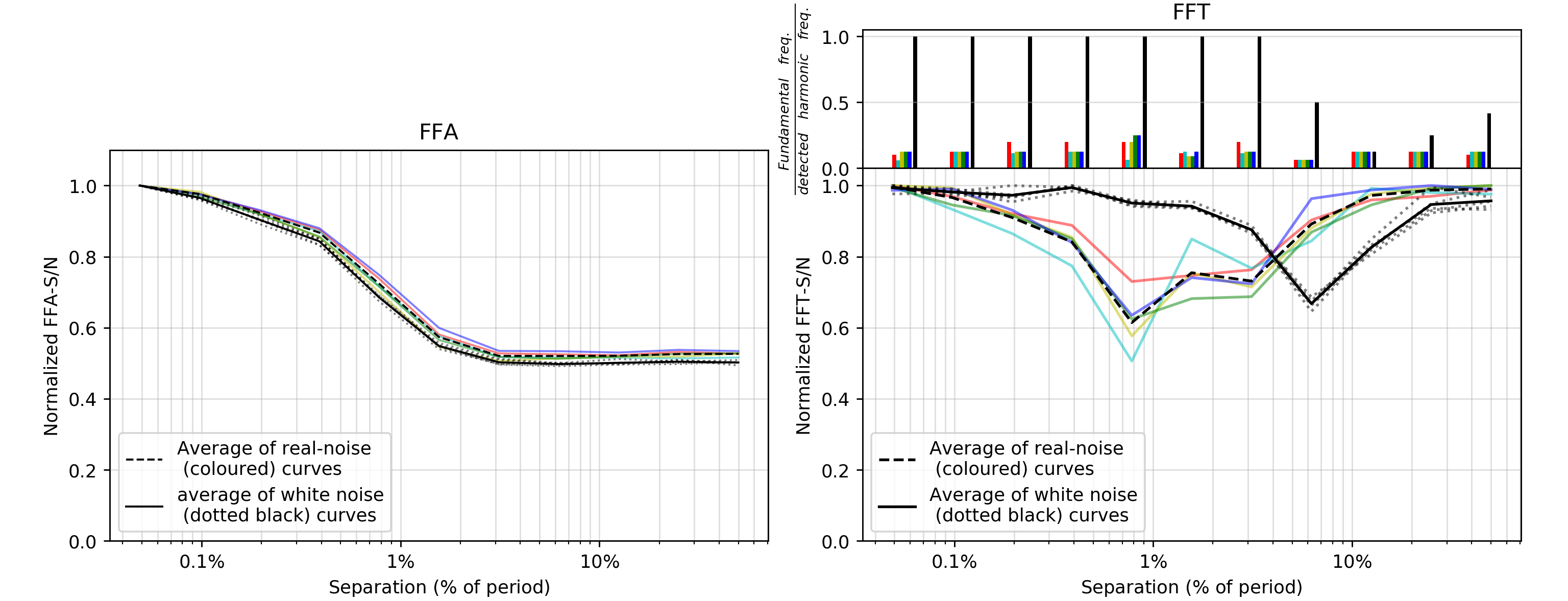}
    \caption{Variation of detection-S/N with separation between components of a double component profile (FWHM of single component is 0.5\% of the period), at period 2s. Each curve represents the results in a particular noise case and is individually normalised to unit maximum. Faint coloured curves are representing results from real GMRT-noise files, while dotted black curves are for white noise cases. The average of real noise cases is plotted as dashed black curve, and average of white noise cases is the solid black curve. In FFA-search, we see a gradual fall in detection significance as separation between components increases. At the full separation, detection S/N of FFA-search fall to half of the initial S/N in both real and white noise conditions. In FFT-search, S/N first decreases and then recovers with increasing separation between two profile components. The dips in FFT-S/N for real noise and white noise cases are happening at different separations. We have the ratio of fundamental frequency and detected harmonic frequency plotted against separation between profile components for each noise case in the upper panel of the FFT plot.}
    \label{fig:snr_separation}
\end{figure}
\begin{figure}[p]
    \centering
    \includegraphics[height=6.8 cm, width=18.5 cm]{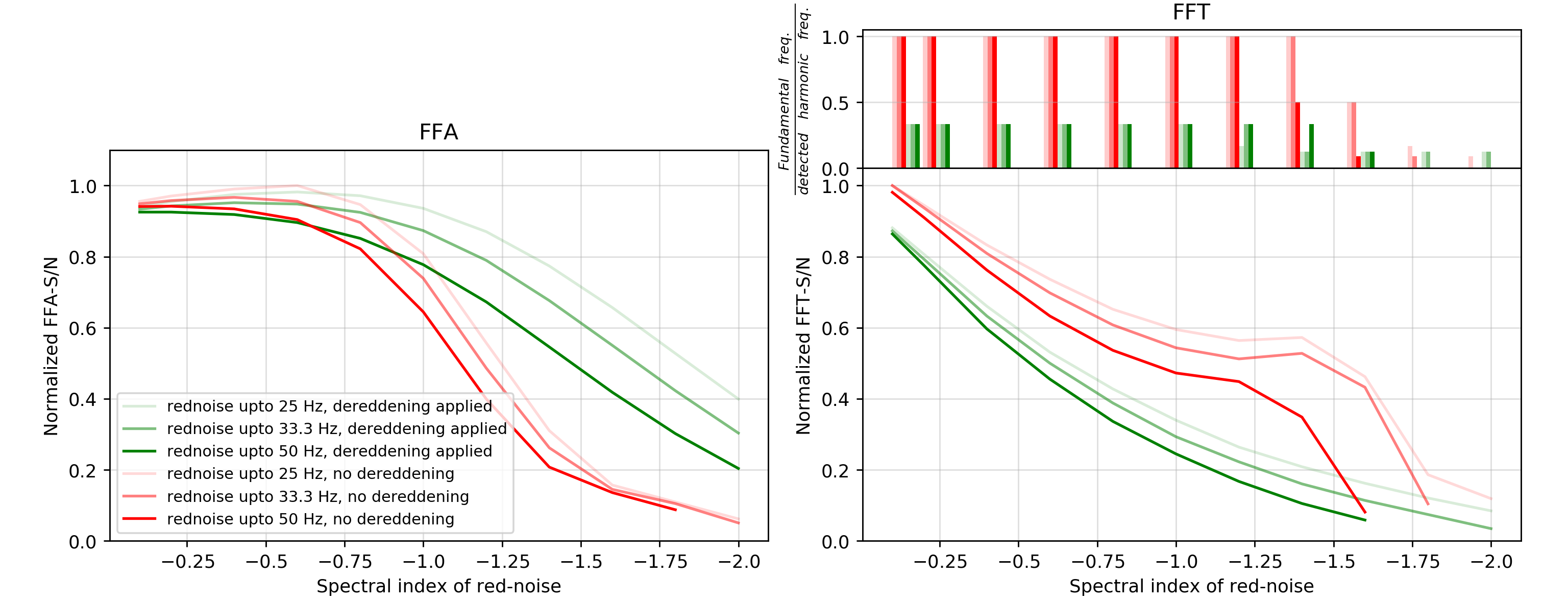}
    \caption{Variation of detection S/N with various realisations of red noise for pulsar at period of 8.5 s. In these plots red curves represent the results of search without any de-reddening process, while green curves are for the search results after de-reddening. We use running median subtraction in time domain for FFA-search, and PRESTO de-reddening routine \texttt{`rednoise'} for FFT-search. Different shades represent different cross frequencies. We see that detection S/Ns for both the search techniques are falling with either the increasing steepness of red noise or the cross-frequency of red noise. The running median subtraction used in FFA-search reduces the effect of red noise and detection S/Ns are improved, while the PRESTO de-reddening routine \texttt{`rednoise'} results in further decrements in FFT detection S/N. In the upper panel of FFT results, ratio of fundamental frequency and detected harmonic frequency has been plotted against spectral index of red noise. Colour and shade of each bar plot is same as their corresponding S/N versus spectral index curve and indicates a trend of detecting higher harmonics with increasing amplitude of red noise. }
    \label{fig:snr_rednoise}

\end{figure}

\subsection{FFA and FFT search performance as a function of duty-cycle}\label{sec:duty_snr}

The duty-cycle of the pulse is a crucial feature of the pulsar signal illustrating the size of emission cone, and can have significant impact on the search sensitivity. Duty-cycle affects the power distribution among the harmonics in the power spectra and hence can greatly impact the FFT search sensitivity involving limited harmonic summation. In this section, we characterize the effect of duty-cycle on the sensitivity of FFA search and FFT search. A detailed theoretical analysis of the effects of duty-cycle on search sensitivity has been done by \citet{RIPTIDE}. Here, we aim to extend that comparative study in the presence of real GHRSS noise.

We injected Gaussian pulses over a duty-cycles range from 0.1$\%$ to 10.0$\%$ (defined by FWHM) for a period of 2 s in the real GHRSS noise and the white noise files. We sampled the duty-cycle ranges from 0.1\% to 1\% and 1\% to 10\% with 5 trial duty-cycle values each. Based on the understanding from known pulsar population, a longer period restricts the range of possible duty-cycles below 10\% \citep{Posselt_2021}. We performed FFA and FFT searches on these data sets to investigate the variations of detection S/N with duty-cycle for each of the search methods. The running median window parameter \texttt{rmed\_width} in \texttt{RIPTIDE} was set to 1 s to make it sufficiently larger than the on-pulse window for maximum duty-cycle and in \texttt{PRESTO}, \texttt{rednoise} module was applied to the power spectra. Normalisation of curves in this simulation is different from that we used in detection S/N versus period plots. The S/N reported in \texttt{RIPTIDE} and \texttt{PRESTO} search is proportional to $\frac{1}{\sqrt{\delta}}$ times the phase averaged injection S/N, where $\delta$ denotes the duty-cycle of pulsar. So, we normalise each detection S/N by the folded profile S/N (obtained from \texttt{prepfold}, representing injection S/N with the same duty-cycle dependence) at the exact period to mitigate its inherent width dependence. Thus, in Fig. \ref{fig:snr_duty} we show that the fractional S/N retrieved by FFA and FFT search methods, is limited only by the response of the underlying algorithms to the input data. Curves generated for FFA S/N versus duty-cycle and FFT S/N versus duty-cycle in white noise conditions are similar to analytical forms obtained by \citet{RIPTIDE}.

The FFA S/N curve shows uniform response over the duty-cycle range in the presence of white noise. It should be noted that we encounter $\sim$ 7\% loss of  
FFA search S/N in the presence of white noise, which was predicted in \citet{RIPTIDE} and can be attributed to the use of boxcars instead of actual Gaussian pulses as templates. The FFA detection S/N in an average of real GHRSS noise cases is also constant over the duty-cycle range, retrieving up to 70\% of the injected folded profile S/N. In FFT search with incoherent harmonic summing, the detection S/N versus duty-cycle curves for white noise cases show an expected increase in limited harmonic space \citep[up to 8 harmonics used in the GHRSS search,][]{GHRSS1}. 
However, the curves for real noise cases decrease marginally with increasing duty-cycle. The average curve of the real GHRSS noise cases decreases with increasing duty-cycle and stays below 20\% of the injected folded profile S/N. The overall degradation of FFT response in real noise cases can also be seen in Fig. \ref{fig:snr_period}, where FFT S/N reduces for longer periods.
In the upper panel of FFT results, we have the ratio of fundamental frequency and detected harmonic frequency (inverse of detected harmonic) versus duty-cycle. Colours of S/N versus duty-cycle and bar plots are same for a given real noise case, and black bar plot is for average of white noise cases. In real noise cases, we see a trend of detecting signals at lower harmonic frequencies as the duty-cycle increases. The degradation of S/N can be explained as an effect of red noise present in real telescope data. With increasing duty-cycle, the fraction of total power in lower harmonics increases and hence the FFT search followed by 8 harmonic summation gets more signal power, resulting in increased significance of FFT detection. 
But this is true only for white noise cases. 
For large duty-cycles in cases of real noise, all the significant harmonics are inside the red noise dominated region and get significantly dampened by spectral whitening, restricting the improvement of detection S/N with increasing duty-cycle.
From this comparison, we can conclude that FFA search has a uniform response for all the duty-cycles both in real noise as well as white noise conditions. FFT search has good sensitivity for larger duty-cycles in white noise, but in the GHRSS noise conditions and for the pulse period of 2 s, FFT detection S/N does not improve with increasing duty-cycle. 

\subsection{Performance of FFA and FFT based search for pulsars with multiple profile components}

The complexity of the pulse profile, accounting for features like the number of components and separation between components, is an important characteristic of the pulsar signal. These features decide the power distribution in the Fourier domain, affecting the FFT detection S/Ns. The folded profile S/N reported by \texttt{RIPTIDE}, is only able to count the power in one of the profile components (within the boxcar filters, ranging up to 20\% duty-cycle), if they are fully separated. Hence, FFA and FFT search sensitivity can be differently impacted by component separations.
Here we aim to find out the effect of separation between two components of the pulse profile on the sensitivity of FFA and FFT search methods.

We injected pulsars with two components, each having Gaussian shape at a period of 2 s in real GHRSS noise and white noise data, where the width (FWHM) of a single component is $0.5\%$ of the pulse period and the separation between these components is logarithmically spaced between 0.05$\%$ to 50.0$\%$ of the period. We considered both the components with similar intensity. 
We kept injection phased-averaged S/N constant for all separations. We performed FFA and FFT (up to 8 harmonics) searches to produce 10 curves (5 real noises and 5 white noises) for each search method. \texttt{RIPTIDE} parameter \texttt{rmed\_width} was again set to 1 s. These curves of detection S/N versus separation between components are individually normalised to unit maximum in order to get them on the same level for comparison.  Fig. \ref{fig:snr_separation} shows the results of FFA and FFT searches as a function of component separation. 

In the case of FFA search, the curves in Fig. \ref{fig:snr_separation} start from $0.05\%$ separation and the S/N decreases with increasing separation. For separation less than 3$\%$ of the pulse period (i.e. components are not fully separated), duty-cycle of the resulting profile increases with the separation between components. With increase in effective width of pulse, S/N reported by FFA search continues to fall. Once the components get fully separated at $3\%$ separation, detection S/N remains at 0.5. This can be explained using the combined effects of duty-cycle on the folded profile S/N and the process of matched filtering with a single boxcar.
\texttt{RIPTIDE} convolves the folded profiles only with boxcar matched filters for pulse detection. Therefore, for widely separated components, only the power of one component can be collected to derive a detection S/N. 
The trend of FFA-S/N is identical in white noise and real noise cases. In the case of FFT search, FFT-S/N trends are different for real noise and white noise conditions. In the white noise case, S/N remains constant up to 3$\%$ separation, after which it drops, and recovers at 50$\%$ separation. We have a similar decrease in the detection S/N for the real noise cases but at different separation. These dips in FFT-S/N versus separation plots are caused by the modulations introduced in the Fourier domain by the double peak of periodic signal. The bar plot (ratio of fundamental-frequency and detected-harmonic-frequency versus separation in the upper panel of FFT results) shows that in the presence of white noise (black bars), we detect signals at higher harmonics only around the separations where we see a dip in S/N. But in majority of the real noise cases (coloured bars), we are always detecting signals at higher harmonics due to the presence of red noise. At the separation where the dip in S/N is seen, a small deviation from the trend in detected harmonics can also be noticed. The reason behind the dip happening at lower separation values in real noise is the presence of red noise. The rednoise restricts us from detecting signals in the fundamental and the first few harmonics, since they are buried inside the red noise dominated region. The `\texttt{rednoise}' routine from \texttt{PRESTO} used for spectral whitening, also introduces another modulation in the harmonic power distribution. Due to this modulation, the harmonic power distribution for a real noise case becomes very different from that of white noise. Since the modulation caused by spectral whitening depends on the rednoise parameters, the separation corresponding to the dip in the S/N also depends on the rednoise parameters. From this comparison, we can conclude that neither of the two search methods (FFA and FFT) has uniform response with separation between two components (or complexity) of the profile. However, sensitivity reduces to a steady 50\% level for well separated components in the FFA search.

\subsection{Performance of FFA and FFT search in various red noise conditions}
In the previous sections, we see that FFT search is severely affected by the presence of red noise in real GHRSS noise conditions. In this section, we inspect the behavior of FFA-search and FFT-search methods in different simulated red noise conditions. 
Red noise is a steep spectral noise, dominant only at lower frequencies in the spectra of real telescope noise. This noise spectra becomes flat (white-noise) at sufficiently high frequencies. Such a combination of red noise and white noise can be characterized by two parameters, the first is cross-frequency, which determines the frequency span of the red noise. This parameter decides the frequency up to which red noise will be dominant. The second parameter is the spectral index or the steepness of the red noise. The cross-frequency combined with the spectral index of the red noise decides the amplitude of the red noise at all frequencies. \citet{timmer} suggested a way to generate red noise, which is to scale the real and imaginary parts of the samples of white noise spectra by the square root of a power-law function.  Following this method, we used spectra of white-noise time series $S(\nu)$ (both real and imaginary) and modified it according to the following expression to get spectra of real telescope like (red noise + white noise) noise $S'(\nu)$,
\begin{equation}
S'{(\nu)}=S(\nu)\times [(\frac{\nu}{\nu_0})^{-\alpha}+1]^{\frac{1}{2}}
\end{equation}
Here $\nu_0$ is the cross-frequency and $\alpha$ is the magnitude of spectral index of red noise. The inverse Fourier transform of $S'(\nu)$  gives the time-series with baseline and rms (Root Mean Square) variations caused by added red noise. We fitted the spectra of 5 real GMRT noise cases with the above expression and found that red noise parameters vary over a large range of values. The cross-frequency ranges between 12 Hz and 60 Hz and the spectral index varies over a range of $-$1.6 to $-$2.3 for these 5 real noise cases. We derive the average value of cross-frequency for GHRSS noise conditions as 32 Hz and the average spectral index as $-$1.9.  We generated a number of red noise injected time series using different cross-frequency (25 Hz, 33.33 Hz, and 50 Hz) and spectral indices ranging from $-$0.1 to $-$2.0 (in 11 regularly separated steps, resulting in a total of 33 simulations). We did not go beyond a spectral index of $-$2.0 in our simulations as signal was not detected in very steep red noise conditions. Then we injected pulses with a periodicity of 8.5 s and a duty-cycle of $1\%$. We used 8.5 s period (i.e. a known pulsar period) to see the effect of red noise and the efficiency of de-reddening schemes for longer period pulsars, which are mostly affected by red-noise. 

We performed FFA and FFT-search on the final time-series data with and without de-reddening. FFA and FFT searches use different methods of de-reddening. Running median subtraction with \texttt{rmed\_width} of 2.0 s was used for FFA-search, whereas FFT-search uses \texttt{PRESTO} based frequency domain spectral whitening (\texttt{rednoise}). For FFT-search, we used flo=0.5Hz in \texttt{accelsearch} to avoid searching in very small frequencies having large red noise amplitudes, which is not relevant for the injected period of 8.5 s.
Fig. \ref{fig:snr_rednoise} shows the results of FFA search and FFT-search with simulated red noise conditions.
FFA and FFT S/Ns are normalised by the maximum S/N we get across all the red noise realizations in the respective search methods. FFA detection S/Ns curves are flat up to spectral index $-$0.6 for each of the cross-frequency choices, but we see significant decrements in detection S/Ns for spectral indices steeper than $-$0.6. Due to the increased amplitude of red noise, detection S/N falls with increasing cross-frequency. The effect of the steepness of red noise becomes more prominent at higher cross-frequencies. In FFA search, running median subtraction improves the detection S/Ns and makes the FFA-S/N versus spectral index curves flatter. In the FFT search results, we see that FFT-S/N falls rapidly with increasing amplitude of the spectral index of red noise. Increasing the cross-frequency results in diminishing the FFT detection S/N, as we have seen for FFA. Moreover, de-reddening of power spectra (by \texttt{PRESTO} based \texttt{rednoise}) reduces the FFT detection S/N further, by down scaling the power in lower harmonics in the process of flattening the power spectra. The bar plots of the ratio of fundamental frequency and detected harmonic frequency (in the upper panel of FFT results) show a trend of detecting higher and higher harmonics with increasing amplitude of red noise. In Section \ref{sec:dereddening}, we further discuss various possible strategies for red noise mitigation. Results from these comparative studies encourage us to use FFA-search with running median subtraction for long period pulsars in cases where the instrumental red noise contribution is non-negligible.

\subsection{Comparison of red noise mitigation techniques}\label{sec:dereddening}

Section \ref{sec:period_snr} describes the role of red noise in degrading the sensitivity of periodicity searches for long period pulsars. The effect of red noise can be reduced in two ways, (1) in the time-domain by a running median subtraction with window size being matched to the timescale of baseline fluctuations; (2) in the frequency-domain by scaling down the excess power in the low modulation frequencies of the power spectra.    
The commonly used pulsar search softwares, \texttt{PRESTO} and \texttt{SIGPROC} employ such frequency-domain red noise removal routines. We use \texttt{PRESTO} based spectral whitening routine \texttt{rednoise} for the frequency-domain de-reddening.

In order to study the effectiveness of these two methods for the cases of FFT and FFA based search in the presence of real GHRSS noise, we used the simulated pulsars (with  real GHRSS noise)  for a range of periods (as described in Section \ref{sec:period_snr}) and duty-cycles (as described in Section \ref{sec:duty_snr}). We made four combinations of search methods and de-reddening routines: FFA search with running median subtraction, FFT search with running median subtraction, FFA search with frequency domain spectral whitening and FFT search with frequency domain spectral whitening. In the case of FFA search with running median subtraction, we first did a running median subtraction on the time series, and then we performed FFA search on the de-reddened time series. For FFT search with running median subtraction, we performed FFT search (using \texttt{accelsearch} of \texttt{PRESTO}) on the spectra of running median subtracted time series. We performed FFA search on the time series obtained by the inverse Fourier transform of the spectra that had undergone frequency domain spectral whitening (using \texttt{rednoise} routine of \texttt{PRESTO}), for the case of FFA search with frequency domain spectral whitening. FFT search was performed on the spectra of time series after doing frequency domain spectral whitening (using \texttt{rednoise} routine of \texttt{PRESTO}) for the case of FFT search with frequency domain spectral whitening.

\begin{figure}[h]
    \centering
    \includegraphics[height=15 cm, width=18 cm]{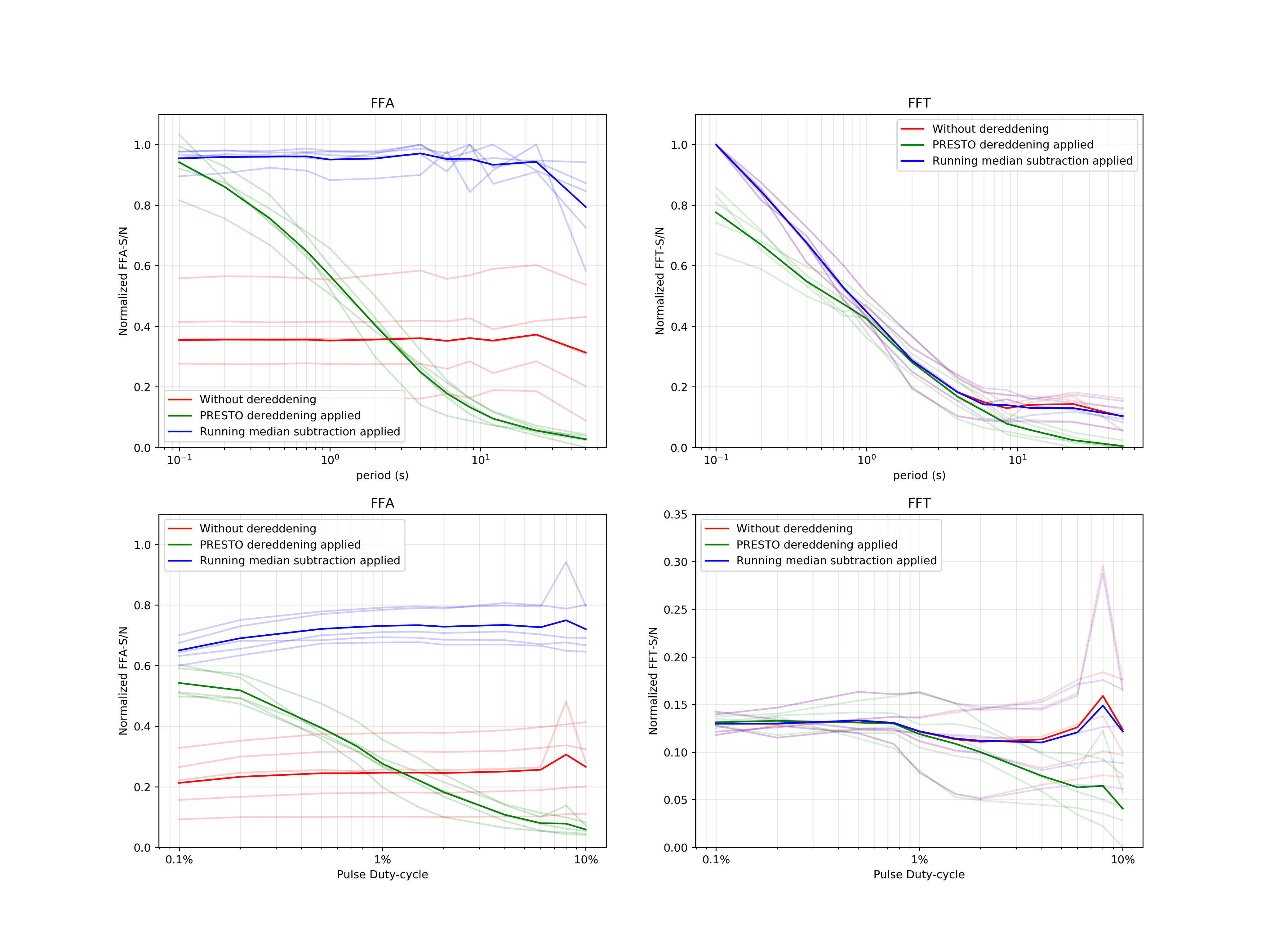}
    \caption{Performance of different de-reddening routines for a range of periods and duty-cycles. The upper panel shows the variation over a range of periods (with 1\% duty-cycle), while the lower panel shows the variation with duty-cycle (2~s period). The left panels are FFA results, while the right panels are results from FFT search. Blue curves represent the behaviour of search S/N when we are using time-domain running median subtraction, red-curves show the S/N when no de-reddening routine is used and green curves represent the detection S/N, after using \texttt{PRESTO}'s frequency-domain de-reddening routine. The improved performances  with time-domain running median subtraction followed by FFA search for long period pulsars, irrespective of their duty-cycle, are evident here.}
    \label{fig:dereddening}
\end{figure}

Fig. \ref{fig:dereddening} shows the comparison of these red noise removal routines. The left panels show the results of FFA search and the right panels show the results from FFT search. Normalisation of plots in upper panel and lower panel are different. The curves for S/N versus period for a specific noise realization are normalised (similar to Section \ref{sec:period_snr}) by the maximum detection S/N obtained among all the periods after de-reddening by running median subtraction (which recovers maximum S/Ns). In the S/N versus duty-cycle plots, we use the folded profile S/N (folded at nominal period and DM) as a metric of maximum recoverable S/N for that simulated pulsar data. The plots for detection S/N versus duty-cycle (lower panels) are normalised by the folded profile S/N of the respective pulsar data (similar to Section \ref{sec:duty_snr}). Hence, the data points represent the fraction of the maximum recoverable S/N.

In the S/N versus period curves, we see that without any de-reddening, the detection S/Ns from FFA search is uniform over the period range. 
However, a change of red noise amplitude for different noise realizations introduces a variation of recovered S/N by a factor of $\sim$5. The FFA search on running median subtracted time-series (blue curves) recovers maximum S/N, whereas whitened power spectra (by scaling the power at low modulation frequencies), provides good improvement for small periods, but the detection S/N falls rapidly with increasing period. This decrease in S/N is likely due to down-scaling of the power at low frequencies having severe effects on long period signals. 
In the S/N versus period curves from FFT search, S/Ns after running median subtraction are identical to the ones without red noise removal, and both are falling with increasing period. Even though running median subtraction of de-dispersed time-series reduces the median and rms at the lower frequency end of the power spectra, the higher median and rms at nearby frequencies suppress the detection significance. De-reddening in the frequency-domain further decreases the FFT detection S/Ns. 

FFA detection S/N is uniform over the range of duty-cycle and with de-reddening by running median subtraction, almost a 3 fold improvement in the S/N is seen compared to the cases of no de-reddening. However, in the case of frequency-domain de-reddening, we see improvement in S/N for smaller duty-cycles, but S/N degrades for larger duty-cycles, likely resulting from down-scaling of the power at low frequencies. With increasing duty-cycle, the fraction of total power in lower harmonics increases and hence down-scaling power in those harmonics reduces detection significance towards the periodic signal. In FFT S/N versus duty-cycle, we see almost constant FFT S/Ns over the duty-cycles for the cases of no de-reddening and de-reddening by running median subtraction. Further reduction in detection S/N is seen for frequency-domain de-reddened data,  which is similar to the behavior seen in the FFT plot of Fig. \ref{fig:snr_duty}. 

From this comparison, we see that in the presence of GHRSS noise for longer periods and all duty-cycles, FFA search with running median subtraction is working more efficiently than de-reddening of power spectra followed by periodicity search either by FFA or FFT. 

\section{FFA search pipeline for GHRSS survey} \label{sec:processing}
The GHRSS survey is an off-Galactic plane survey ($\left|b\right|>5$ degrees) conducted with the GMRT in two phases. The first phase was with the legacy GMRT system (data from GMRT Software Back-end (GSB), \citet{GSB}), with a bandwidth of 33 MHz over 306-338 MHz (\cite{GHRSS2}, \cite{GHRSS1}) resulting in 13 discoveries, and the second phase is with the upgraded GMRT (uGMRT, data from GMRT Wideband Back-end(GWB), \citet{GWB}), having a bandwidth of 200 MHz over 300-500 MHz (Bhattacharyya et al. in prep) resulting in 15 discoveries (including 3 Rotating Radio Transients) till now. Time and frequency resolutions for phase-I were 30.72 or 61.44 $\mu s$ and 32.552 or 16.276 kHz based on the Galactic coverage. In phase-II, the GHRSS survey has a time and frequency resolution of 81.92 $\mu s$ and 48.828 kHz, respectively. 

We have implemented an FFA (\texttt{RIPTIDE}) based search pipeline\footnote{ https://github.com/GHRSS/ffapipe} for the GHRSS survey. According to the ATNF pulsar catalog, the long period (P $>$ 1s) pulsar population in the off-Galactic plane is $\sim$ $28\%$. This estimate of long period pulsar population could be a lower limit as the majority of the current pulsar population was discovered using FFT search with shorter integration times. Before FFA search processing, GHRSS had only 10\% of its discoveries with P $>$ 1s. Currently, after a year of FFA processing, the long period pulsar population in the GHRSS survey increases to 20\%. In Section \ref{sec:discoveries}, we report some of these new discoveries.\\
With the FFA pipeline, we are concurrently processing both the archival as well as new observations of the GHRSS survey. The archival data was searched for pulsars using conventional FFT based periodicity search. The major part of the archival data comes from phase-I of the survey, where the bandwidth is 33 MHz. The FFT search pipeline for phase-I was described in \citet{GHRSS1}.   The new data coming from this survey is undergoing both FFT search (\texttt{PRESTO} based pipeline) and FFA search (\texttt{RIPTIDE} based pipeline) in order to obtain optimal sensitivity for all periods and duty-cycles.

\begin{figure}[ht!]
    \centering
    \includegraphics[height=10 cm, width=18 cm]{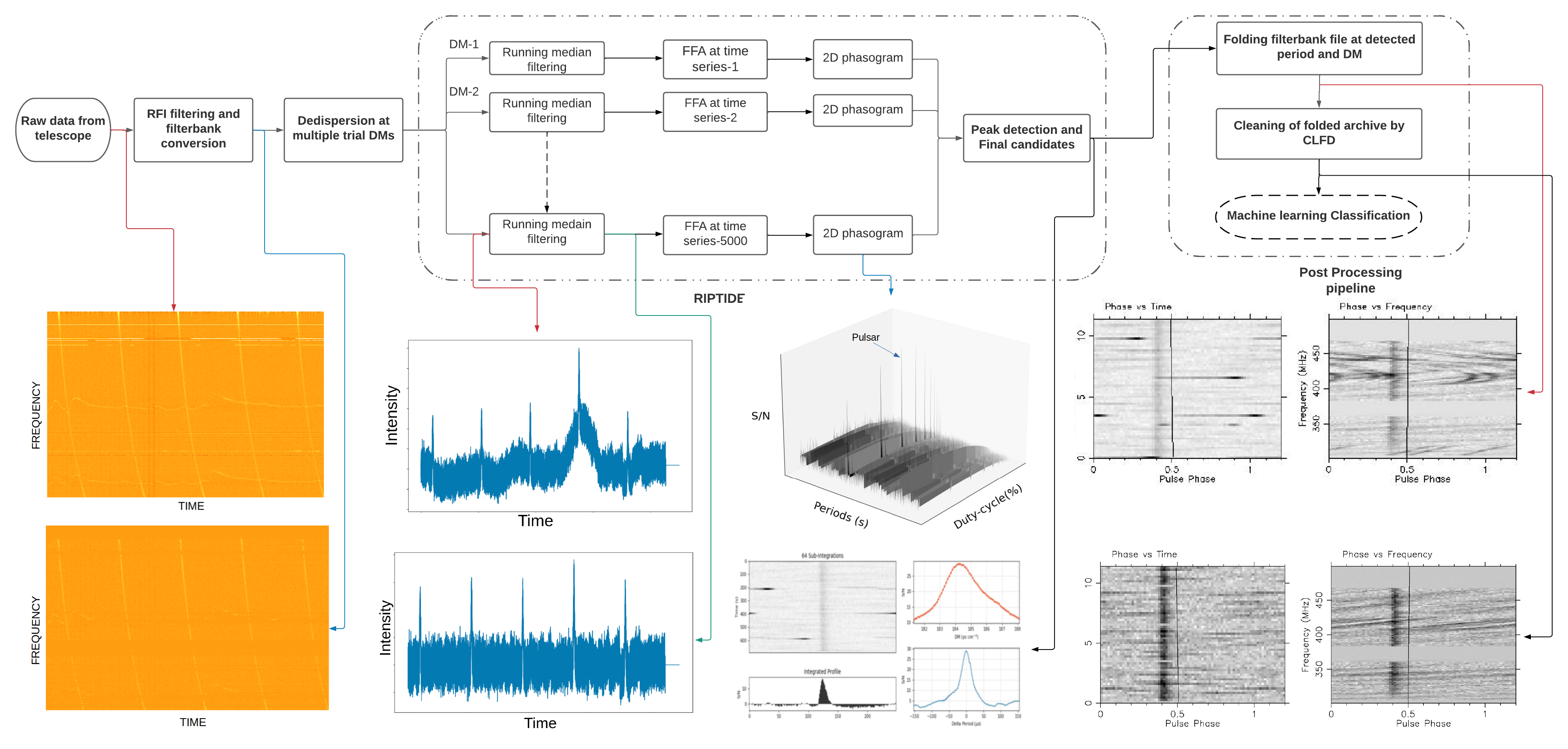}
    \caption{ Block diagram of the FFA search pipeline implemented for the GHRSS survey along with visualisation of signals at various stages of the pipeline.}
    \label{fig:block_diagram}

\end{figure}

In Fig. \ref{fig:block_diagram}, the FFA search pipeline for the GHRSS survey is illustrated. The first step of the processing deals with the cleaning of \texttt{filterbank} data and de-dispersion at a number of trial DM values. De-dispersion is the process of adding signals from all the frequency channels after correcting for the frequency-dependent time delay introduced by the interstellar medium. This produces a frequency averaged dedispersed time series.  After this, the periodicity search routine of the pipeline performs running median subtraction on dedispersed time-series to reduce red noise contamination and then carries out FFA search and produces final optimized candidates. The candidate list contains information about DM, period, duty-cycle, and S/N. The \texttt{filterbank} data is then folded at these detected periods and DM values in the post-processing stage. These folded data-cubes are also cleaned by \texttt{CLFD}\footnote{https://github.com/v-morello/clfd}\citep{Morello_2018} and final cleaned data-cubes are provided to the machine learning based classifier used in the GHRSS survey \citep{Lyon_2016}. The machine learning classifier working on the folded data-cubes separates pulsar signals from non-pulsar candidates.

\subsection{RFI filtering and dedispersion}

In the first step of the survey processing, we filter out  RFIs from the time-frequency \texttt{filterbank} data. The bright spectral and temporal features can be seen in the raw dynamic spectrum presented in Fig. \ref{fig:block_diagram}. We use \texttt{gptool} (Chowdhury et al. in prep ) for data cleaning. In the first step, \texttt{gptool} calculates the system bandpass and normalizes the spectra of the data block by the system bandpass. This makes the identification of spectral line RFIs and replacing them by the mean value much easier. After correcting for the system bandpass, simple thresholding is used to flag spikes above 3$\sigma$ in the spectra. Following the flagging of spectral line RFIs, a frequency averaged time series is produced. A histogram of these time samples is used to calculate the FWHM and the mean of the distribution, which are then used to flag the outliers above 3$\sigma$ threshold in the time series. The final cleaned filterbank data (as seen in Fig. \ref{fig:block_diagram}) is written to disk for further processing. We also use \texttt{zerodm} subtraction routine available in \texttt{SIGPROC} distribution to remove broadband RFI.  

We search in the DM range of 1 $pc~cm^{-3}$ to 500 $pc~cm^{-3}$ with DM-step of 0.1 $pc~cm^{-3}$, resulting in $\sim 5000$ trial DM values. The DM range is decided to keep the DM limit greater than the maximum DM encountered in the GHRSS sky having $\left|b\right|>5$ degrees, according to YMW16 electron-density model\footnote{https://www.atnf.csiro.au/research/pulsar/ymw16/}\citep{YMW16}. \texttt{PRESTO's} \texttt{prepsubband} routine is used to generate dedispersed time series for these 5000 trial DM values down-sampled to $\sim$ 1 ms. This down-sampling helps us to speed-up the search process. This also limits the duty-cycle resolution (smallest duty-cycle being searched) of the search. The lowest duty-cycle resolution for the smallest period being searched (100 ms) is 1\%, which is adequate as smaller period pulsars are expected to have larger duty-cycles\citep{Posselt_2021}. 

\begin{figure}[p]
    \centering
    \includegraphics[height=16 cm, width=18 cm]{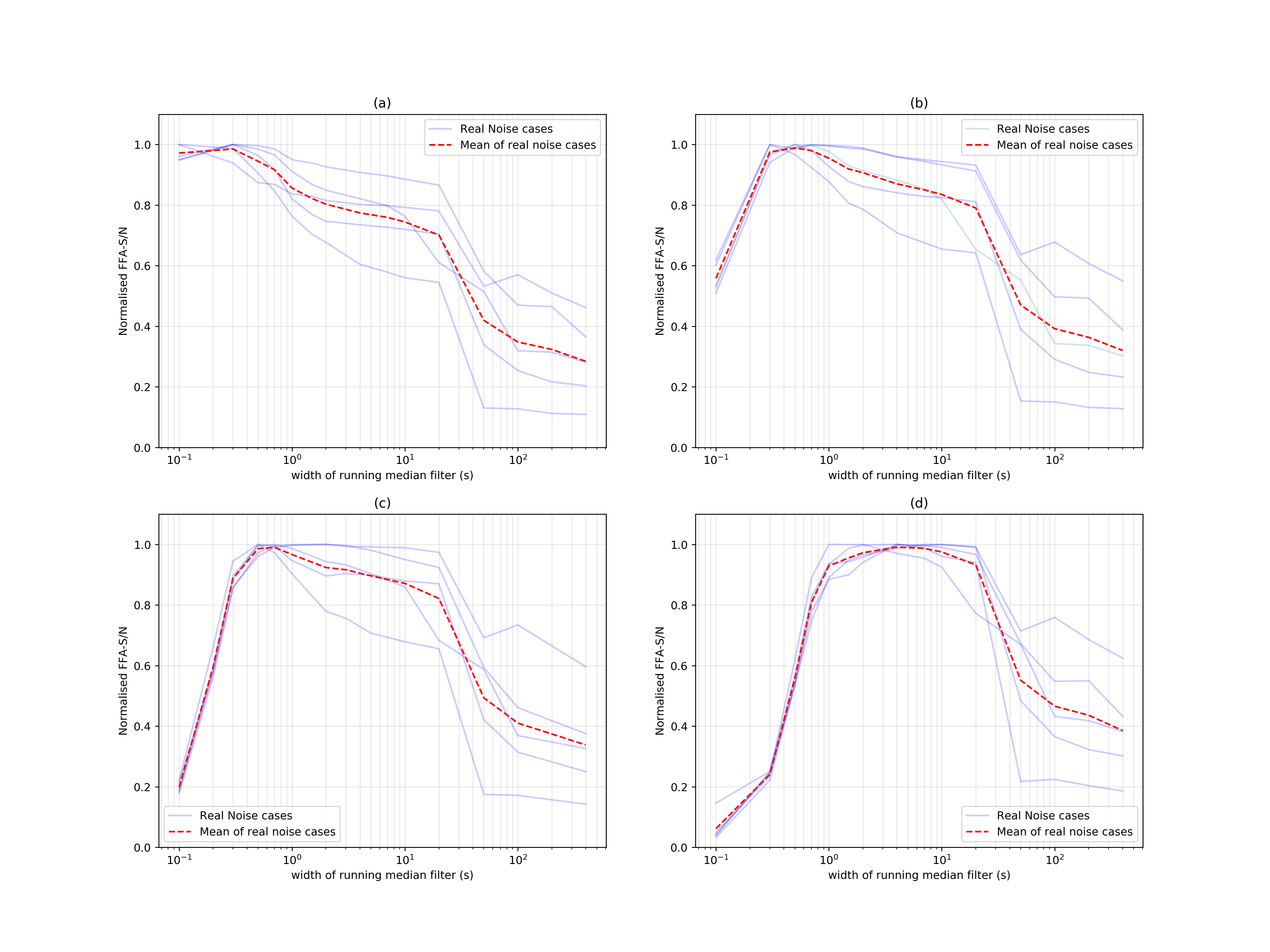}
    \caption{FFA detection S/N versus running median window width for four representative candidates \textbf{from} the four period ranges used in the GHRSS pipeline. In these plots, faint blue curves are for various real noise cases and the dashed red curve is the mean of real noise cases. For the period range 0.1$-$0.5 s (plot (a)), we used period 0.4 s and duty-cycle 3.1 $\%$ as a representative candidate, for range 0.5$-$2.0 s (plot(b)), the representative candidate has period 1.5s and duty-cycle 3.1$\%$ , for period range 2.0$-$10.0 s (plot(c)), the pulsar has period 8.5 s and duty-cycle 1$\%$ and for period range 10$-$100 s (plot (d)), it is with period of 23.5 s and duty-cycle of 1$\%$. In these plots, S/N decreases at very small values (signal is getting subtracted by running median subtraction) as well as very large values (red noise is not getting removed) of running median window. We choose the running median window size for these period ranges by taking into account the behavior of FFA S/N with the window width and the maximum possible duty-cycles that we want to detect in each of the search period ranges.}
    \label{fig:rmed_width}
\end{figure}

\subsection{Optimal running median window for de-reddening}\label{sec:r_med}

Running median subtraction has the property that it leaves those structures unharmed which have a size less than half of the median window size. This effectively removes all the baseline variations having time-scales greater than the median window. But, it can remove pulse signal if width of window is less than double of the on-pulse duration. Thus, the running median window size should be kept at least double of the largest on-pulse duration we are searching for. There is an additional constraint on the searchable on-pulse duration from the time scales of dominant baseline variations introduced by red noise. We can not search for the pulses which have a duration similar to baseline fluctuations. So, we must decide the running median window size considering the maximum desired on-pulse duration and the red noise conditions.

In Fig. \ref{fig:rmed_width}, we plot variations of FFA S/N with running median width for different folding periods. 
Like earlier simulations, we used the same five GHRSS noise files and injected pulses at four different periods, resulting in 20 simulated pulsar data. These four periods (0.4~s, 1.5~s, 8.5~s and 23.5~s) represent the four period ranges (0.1~s$-$0.5~s, 0.5~s$-$2.0~s, 2.0~s$-$10.0~s and 10.0~s$-$100.0~s) used in the GHRSS FFA pipeline. We configure the GHRSS search pipeline to use separate running median width, search parameters, and candidate optimization parameters for these period ranges. Considering the median duty-cycle of pulsars in ATNF Pulsar Catalog\footnote{https://www.atnf.csiro.au/research/pulsar/psrcat/} we used duty-cycle 3.1$\%$ for periods less than 2~s and $1\%$ for periods greater than 2~s in these simulations. S/N versus width of running median window curves are individually normalised to unit maximum for each real-noise case.

The upper left panel in Fig. \ref{fig:rmed_width} (a) is for folding period 0.4~s and duty-cycle 3.1$\%$. This plot has a peak in S/N at a running median window width of around 0.3~s and falls rapidly with increasing the window widths. The upper right panel (b) is for period 1.5~s and duty-cycle 3.1$\%$. The FFA S/N in this plot peaks at 0.5~s and is falls at either side of the running median window size axis. The decrease at the beginning is due to subtracting some part of the pulse itself due to the smaller size of the window. The fall on the right side is due to residual baseline fluctuations, which are not removed due to the larger size of the running median window. The lower panels (c and d) are for periods 8.5~s and 23.5~s and duty-cycle 1$\%$ respectively. S/Ns in these plots also fall on either side of the running median window size axis and peak at 0.7 and 3~s respectively. If the time-scales of the on-pulse window are similar to the time-scales of dominant baseline variations, then we can not distinguish between a pulsating signal and variations due to red noise. For GHRSS noise, the time-scales of typical baseline fluctuation are of the order of seconds. So, searching for pulses of width greater than a few seconds is not beneficial, hence the maximum running median width can be kept around 4$-$5~s. We keep the running window width at 0.5~s for the search period range of 0.1$-$0.5~s, allowing for duty-cycles up to 50$\%$ for the maximum period in the range. The window width is kept at 1.0~s for the period range of 0.5$-$2.0~s allowing for duty-cycle 25$\%$ for a 2~s period. For the period range 2~s$-$10~s, we keep the running median window width at 2~s allowing 10$\%$ duty-cycle for a 10 s period and more for other periods in the range. For the period range of 10~s$-$100~s period, the window is kept at 5~s.

\subsection{FFA search and candidate selection using RIPTIDE}

 The efficiency of a search technique depends on the fine-tuning of the search parameters according to the properties of the data being searched and the properties of desired candidates. In the search for pulsars in the GHRSS data over a period range of 0.1~s to 100~s, we create four configurations of periodicity search. These configurations are the following: short (0.1~s$-$0.5s~), medium (0.5~s$-$2.0~s), long (2.0~s$-$10.0~s) and ultra-long (10.0~s$-$100.0~s). Separate search and detection parameters (e.g. bins\_min, bins\_max, and running median window size) are used for these four search configurations. We notice that rednoise in phase-I data is less severe and should not affect the FFT search performance for the period range corresponding to short configuration (0.1~s$-$0.5~s), hence we limit FFA search over a 0.5$-$100 s period range for phase-I data.
 
 \texttt{RIPTIDE} takes a set of dedispersed time series as input and performs FFA search and candidate selection. \texttt{RIPTIDE} outputs a number of data products, including files with parameters of detected candidates and other diagnostic information. The pipeline performs the following major tasks: data-whitening and normalisation; searching over a period range and then matched filtering with a set of boxcars which generates a periodogram, peak detection in the resulting periodogram and then peak clustering. After matched filtering, we get S/Ns of folded profiles for detection and candidate optimisation.
 For data whitening, \texttt{RIPTIDE} uses running median subtraction (as seen in Section \ref{sec:r_med}). The optimal running median windows are chosen for the four search configurations as described in Section \ref{sec:r_med}. \texttt{RIPTIDE} uses these different values of window size while searching in these different search configurations. After subtracting the running median, the data is normalised to zero mean and unit variance. 
 Then a periodicity search is performed on this normalised time series using folding transform (refer to \cite{RIPTIDE} for more details on these processes). After the search process is over, we have a 2D array of best S/Ns, with dimensions of trial period and trial width (trial boxcar width used for matched filtering). Now, we need to perform peak detection on this 2D array. 
 The non-uniformity in S/Ns introduced by correlated noise (resulting from baseline variations in the folded profiles of candidates) suggests against applying a constant threshold for peak detection. The widest matched filters have optimal response for these baseline variations, providing higher S/Ns. Thus, in the case of a constant threshold, all the candidates with a period greater than a certain value will have S/N above the cut-off. To circumvent this problem, \texttt{RIPTIDE} evaluates the significance of candidates in a local distribution of candidates having similar values of width and period. 
 \texttt{RIPTIDE} separately performs peak detection for each trial width. For each of the trial widths, it takes detection S/Ns as a function of modulation frequency ($\frac{1}{P}$). It divides the modulation frequency range into sub-ranges and computes the median and standard-deviation of S/Ns for each of the sub-ranges. The median values of S/Ns are then fitted as a function of modulation frequency. Now, any candidate with an S/N above 7 times the local RMS from the fitted curve is called a detection. Once all the time-series have been searched for periodic signals and the results have undergone the peak detection algorithm, the peaks from all time-series are grouped into clusters. Clusters that are harmonically related are flagged. Now, for each remaining cluster, \texttt{RIPTIDE} produces a candidate object. 
 
 \subsection{Post-processing}
 The candidates generated by \texttt{RIPTIDE} contain all the necessary information required for classification, except the sub-band versus phase information. Frequency versus phase information is crucial to classify the broad-band nature of the candidates to distinguish between pulsars and non-pulsars. We generate new diagnostic data by folding the filterbank file over 128 sub-bands at the detected period and DM value of the candidate. Folding of a filterbank file generates a folded data cube binned over sub-integration, sub-band, and rotation phase. The aim of the post-processing pipeline is to generate a clean data cube and use machine learning to classify the candidates as pulsars or non-pulsars.

 CLFD \citep{Morello_2018} is used to clean the folded data cubes. CLFD masks the bad sub-integration and sub-bands from the folded data cubes. This significantly improves the S/N of the folded profile along with mitigating artifacts in the sub-integration versus phase plot and sub-band versus phase plot (as seen in Fig. \ref{fig:block_diagram}). This can also improve the efficiency of the machine learning classifier used for the GHRSS survey. GHRSS machine learning pipeline \citep{GHRSS1} is based on \citet{Lyon_2016}, which employes Gaussian-Hellinger Very Fast Decision Tree (GH-VDFT, \citet{VFDT}) classifier for identifying pulsar signals in unbalanced candidate data. The input to the classifier is four features (i.e. mean, standard deviation, kurtosis, and skew) each for  integrated pulse profile and S/N versus DM curve extracted from the folded data-cube. The candidates classified as pulsars by the classifier are then used for human inspection. However, we are in the process of integrating the classifier directly working on the RIPTIDE output based on features like subintegration versus phase, S/N versus period, S/N versus DM, S/N versus width, and the integrated profile of the pulsar.

\section{Results from survey processing} \label{sec:discoveries}

In this paper, we report results from the processing of 1100 deg$^2$ of phase-II GHRSS data and 400 deg$^2$ of phase-I GHRSS data with the FFA search pipeline. We have discovered two new pulsars in the data from phase-II of GHRSS, one of them was also subsequently detected by FFT search, but with reduced S/N. One of these pulsars (J1517$-$31b) with a period of 1.103~s has an unusually narrow duty-cycle and was uniquely detected by the FFA search. The duty-cycle of this pulsar is 0.44\%, which is shorter than the predicted lower limit of 0.77\% for this period \citep{MSPES2016}. Table \ref{table:1} lists the discovery parameters of these two pulsars. PSR J1517$-$31a and J1517$-$31b are seen within the same GHRSS pointing with HPBW (Half Power Beam Width) $\sim$ $\pm$ $32'$. The flux density is estimated from the incoherent beam detection without any primary beam correction (as the pulsar location is not known within the beam). The discovery plots for these pulsars are given in Fig. \ref{fig:J1517a} and \ref{fig:J1517b}. The sub-band versus phase plots were extracted from the folded data-cubes. These new pulsars are being followed up with the uGMRT. The timing solutions with accurate locations of these pulsars will be published soon. One of the GHRSS pulsars discovered in phase-I, PSR J1947$-$43 \citep{GHRSS1} was earlier detected at higher harmonics (7th one) of the true period in FFT search due to the presence of red noise. In re-processing with the FFA pipeline, the true period of the pulsar is corrected from 180.94~ms to 1.266~s. In addition to the two discoveries reported in this paper, we independently discovered PSR J1845$-$40 at a DM of 47.4 pc cm$^{-3}$ and a period of 373.4 ms, from the FFA search, which was discovered by the GBNCC survey \citep{McEwen_2020}. 

\begin{figure}[H]
    \centering
    \includegraphics[height=7.5 cm, width=19 cm]{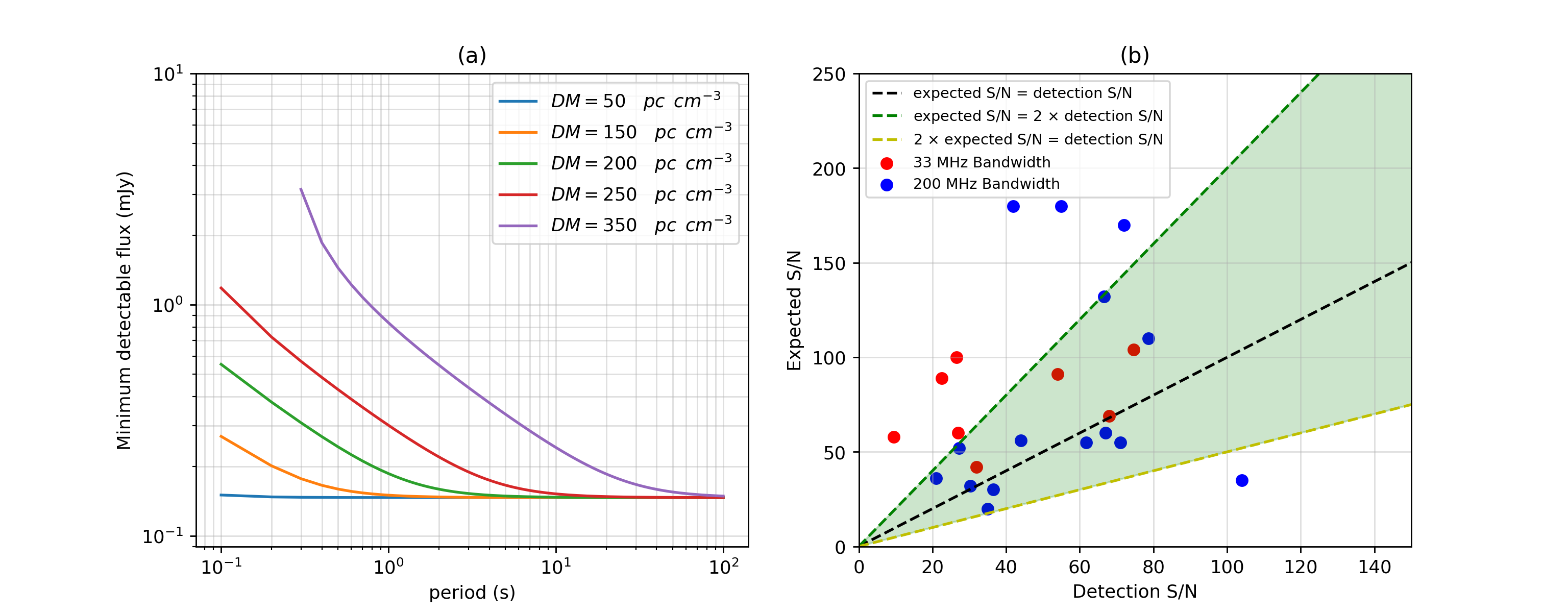}
    \caption{Variation of theoretical minimum detectable flux (at 7$\sigma$ significance) as a function of period and DM (for 1\% duty-cycle) is presented in the plot (a) and a comparison of expected theoretical S/N and detection S/N is presented in the plot (b). Minimum detectable flux increases with increasing DM values and the GHRSS survey is sensitive to find pulsar with flux densities $>$ 0.15 mJy for periods greater than 1~s (for DM $<$ 150 pc cm$^{-3}$). In the plot (b), expected S/N is plotted against actual detection S/N, and the black dashed line represents a 1:1 line (where expected S/N is equal to detection S/N). The shaded part shows the region on the plot where expected S/N differs from detection S/N by a factor of less than two.}
    \label{fig:theoretical_snr}
\end{figure}

The theoretical sensitivity of a survey is a function of the rotation period and DM.
We computed the theoretical minimum detectable flux density ($S_{min}$) using radiometer equation \citep{handbook} considering an efficiency factor of FFA search 0.93 \citep{RIPTIDE} over different periods and DMs for the GHRSS survey at band-3 (300$-$500 MHz) assuming a duty-cycle of 1$\%$. The left panel of Fig. \ref{fig:theoretical_snr} (a) shows the variation of $S_{min}$ over a period range of 0.1 s to 100~s. The plot shows that the theoretical sensitivity of the GHRSS FFA search is 0.15 mJy (7$\sigma$) for periods greater than 1 s and DM less than 150 pc $cm^{-3}$. However, the RFI and temporal variation of the instrumental gain limit the achieved sensitivity.  
In order to compare the sensitivity we achieved with the theoretical sensitivity, we used the detection significance of the re-detected pulsars. We detected 43 known pulsars in 48 pointings from the data processed so far with the FFA pipeline. For 23 out of 48 re-detected pulsars, we were able to calculate expected S/N (using catalog flux density, spectral index, and primary beam shape). For the remaining 25 re-detected pulsars, either flux density was not available at any frequency in the literature or they were too far from the pointing center for a reliable correction for primary beam effect. The right panel of Fig. \ref{fig:theoretical_snr}(b) shows the expected S/N versus detected S/N for these 23 pulsars. Six pulsars from phase-II data have more detection S/N than the expected theoretical S/N (points lying below the black dashed line). This could happen due to uncertainty in estimation of flux density at 400 MHz and/or presence of scintillations. We estimated flux density at 400 MHz from the 1400 MHz flux value given in the ATNF catalog\footnote{ https://www.atnf.csiro.au/research/pulsar/psrcat/}assuming a spectral index of $-$1.5 for these pulsars. The majority of pulsars are lying within the shaded region, indicating up to a factor of two degradation of survey sensitivity. We conclude that the FFA search pipeline of the GHRSS survey is sensitive to find pulsars with 7$\sigma$ $>$ 0.3 mJy for periods greater than 1~s. 

Since in the GHRSS survey we are concurrently processing the data both with FFA and FFT search (with up to 8 harmonic summing) pipelines, we compare the performance of FFA and FFT Search on the 48 re-detected pulsars. We plot FFA-S/N versus FFT-S/N (similar to Fig. 11 \& 12 in \citet{Cameron_2018}) to investigate the general trend. 

\begin{figure}[H]
    \centering
    \includegraphics[height=7.5 cm, width=19 cm]{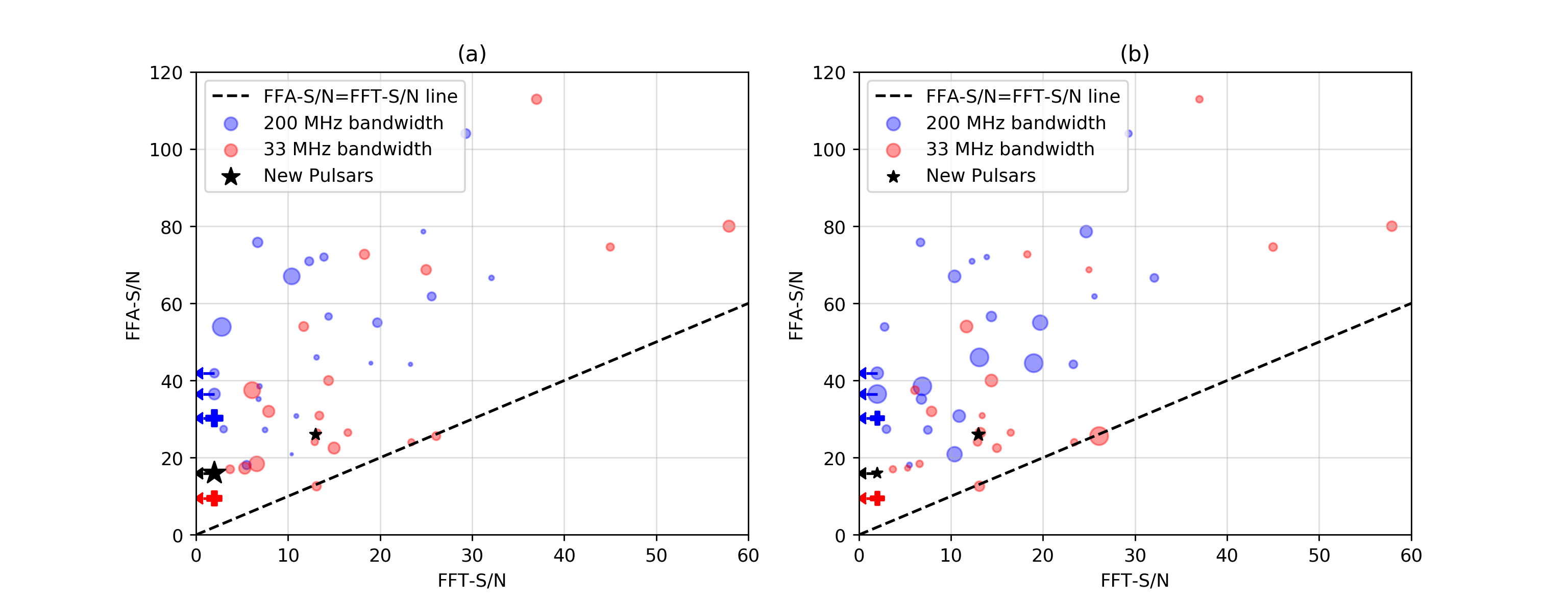}
    \caption{Comparison of FFA-S$/N$ and FFT-S$/$N as a function of period and duty-cycle for the 43 known pulsars re-detected in 48 pointings of the GHRSS survey. In plot (a), size of circles are proportional to period of the pulsar and it's proportional to duty-cycle in plot (b). Symbol '*' represents the new pulsars discovered by FFA search. The black dashed line represents the 1:1 line (where FFA S/N is equal to the FFT S/N). Plot (a) implies that pulsars with longer periods are biased towards giving higher S/N in FFA search. 
    Whereas no such dependence is evident from plot (b) showing a similar comparison for the duty-cycle. Pulsars missed by FFT search with 8 harmonic summing are plotted with 2$\sigma$ upper-limit in FFT S/N. Two of these missed pulsars were detected in FFT search with 32 harmonic summing and are marked as filled plus ('+') signs in the plot.}
    \label{fig:real_pulsars}
\end{figure}

\begin{figure}[H]
    \centering
    \includegraphics[height=7 cm, width=18 cm]{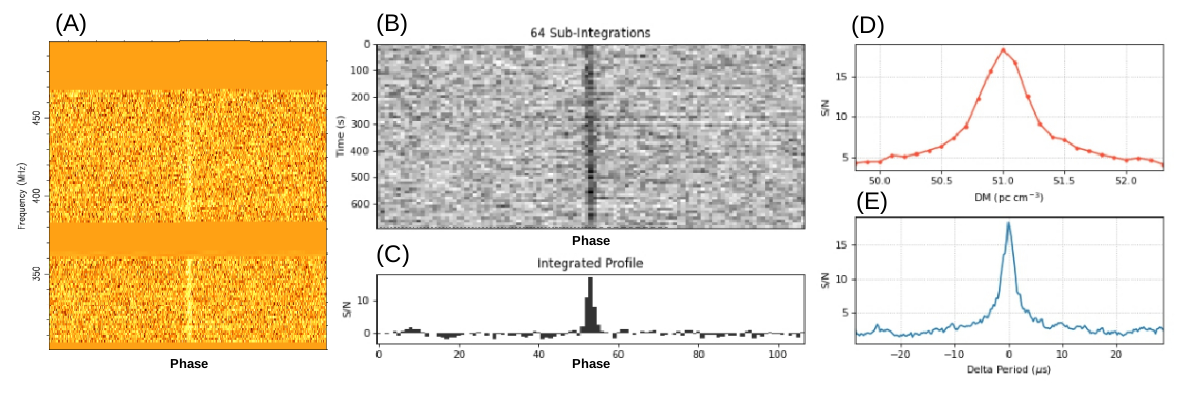}
    \caption{Detection of pulsar J1517$-$31a, discovered in the GHRSS survey. This is a 140.6 ms pulsar having a DM of 51.0 pc cm$^{-3}$. (A) Frequency versus pulse phase plot, obtained from post-processing and added to \texttt{RIPTIDE} output for better visualisation of broadband nature of the signal. (B) Time versus pulse phase plot. (C) Integrated pulse profile given by \texttt{RIPTIDE}. (D) Plot of achieved S/N versus DM, and (E) Plot of the S/N versus pulse period.} 
    \label{fig:J1517a}
\end{figure}

\begin{figure}[H]
    \centering
    \includegraphics[height=7 cm, width=18 cm]{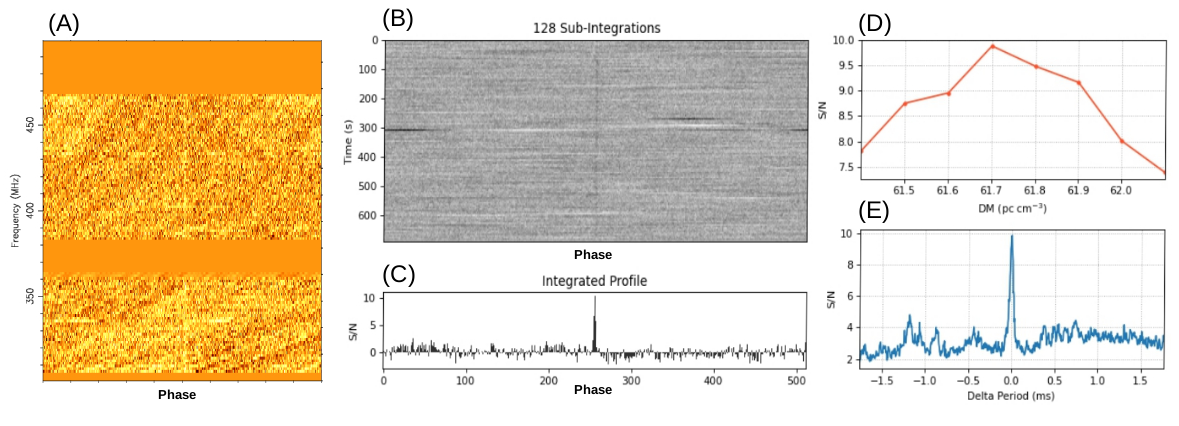}
    \caption{Detection of pulsar J1517$-$31b, discovered in the GHRSS survey. This is a 1103.7 ms pulsar having a DM of 61.7 pc cm$^{-3}$. The panels are the same like Fig. \ref{fig:J1517a}.} 
    \label{fig:J1517b}
\end{figure}

Fig, \ref{fig:real_pulsars} (a) shows the comparison of FFA and FFT S/N for the known pulsars, where the position of a pulsar in this plot is marked by filled circles having a size proportional to the pulsar period. 
The red circles are detections from the phase-I data with 33 MHz bandwidth system and the blue circles are detection from phase-II data with 200 MHz bandwidth system. The black star ('*') symbols are for the newly discovered pulsar reported in this paper. A 1:1 dotted line is added to the plot. The plot shows that pulsars with longer periods are biased towards giving more S/N in FFA search. This result is consistent with the findings of Section \ref{sec:period_snr} (using simulated pulsars in the presence of real noise), where we conclude that FFA search has a uniform response for all periods, while FFT search sensitivity falls with increasing period.
Fig. \ref{fig:real_pulsars}(b) shows the comparison of FFA and FFT S/N, where  the marker (filled circles) size is proportional to the duty-cycle of the pulsar. Even though all the points are above the 1:1 line, we do not see any clear evidence of dependence on duty-cycle. This result is also consistent with the findings of Section \ref{sec:duty_snr}, where we conclude that in the presence of real telescope noise, both FFA and FFT search have uniform response over duty-cycle. Out of these 43 re-detected known pulsars and 2 new pulsars, 5 pulsars were missed by FFT search with 8 harmonic summing (marked by 2$\sigma$ upper-limits of FFT search S/Ns). Among these 5 missed pulsars, 3 are from phase-II data, 1 is from phase-I, and the other one is a new pulsar. We performed FFT search on these pointings with 32 harmonic summation as well. In this search, we detected 2 pulsars (marked by filled '+' sign in fig. \ref{fig:real_pulsars}) out of 5 previously undetected ones with detection significance $\sim$ 2$\sigma$ at fundamental frequency. Three pulsars, including one new discovery, J1517$-$31b were missed by FFT search even with 32 harmonic summing.

\begin{deluxetable*}{cccccccc}
\tablenum{1}
\tablecaption{New pulsars were discovered by FFA search in the GHRSS survey. The discovery parameters like period, DM, duty-cycle and flux densities at 400 MHz (not corrected for primary beam effect) are listed. The detection S/N from FFA and FFT searches (with 8 and 32 harmonic summation) are also added. We found these two pulsars in the same IA beam centered at ($15h17m51.86s, -31d20'40.0''$) and having a Half Power Beam Width (HPBW) of $64'$. One of these two pulsars was detected only in the FFA search.}
\tablewidth{0pt}
\tablehead{
\colhead{Pulsar name} & \colhead{Period } & \colhead{DM} &  \colhead{$S_{400}$} & \colhead{FFA-S/N} & \colhead{FFT-S/N} & \colhead{FFT-S/N} & \colhead{Duty-cycle}\\
\colhead{} & \colhead{(ms)} & \colhead{(pc cm$^{-3}$)} & \colhead{(mJy)} & \colhead{} & \colhead{\texttt{numharm=8}} & \colhead{\texttt{numharm=32}} & \colhead{Based on $W_{50}$} 
}

\startdata
J1517$-$31a & 140.6677(1) & 51.00(2) & 0.6 & 26 & 13 & 22 & 1.5\% \\
J1517$-$31b & 1103.7264(4)& 61.70(6) & 0.4 & 16 & not detected & not detected & 0.44\% \\
\enddata
\end{deluxetable*}
\label{table:1}


\section{Summary} \label{sec:discussion}

In this paper we discuss the possible improvement of the GHRSS survey sensitivity from the FFA search for long period pulsars. 
 Comparison of FFT and FFA search techniques using simulated as well as real pulsars clearly shows that FFA search is providing better sensitivity for long period pulsars in the GHRSS survey. Also, in real GHRSS noise conditions, FFT search sensitivity is not improving for large duty-cycle pulsars due to the presence of red noise. Since FFA is a fully coherent search method, it shows better sensitivity for any period and duty-cycle. For the targeted pulsar population in the GHRSS noise condition, red noise mitigation by running median subtraction in the time-series outperforms the scaling of low frequency signals in power spectra (implemented in \texttt{PRESTO} and \texttt{SIGPROC}). 

From the FFA search of GHRSS data, we re-detected 43 known pulsars, where 4 of these having good S/Ns in the FFA search were missed by the FFT search with 8 harmonic summing. Two of these four missed pulsars appeared in the FFT search with 32 harmonic summing. Also, one of the new pulsars reported in this paper, discovered uniquely in FFA search, was missed by FFT search even with 32 harmonic summing. 
From the comparisons of detection S/N from FFA and FFT searches of known pulsars and new discoveries as well as fitting of red noise parameters, it is apparent that the GHRSS data from phase-I (33 MHz bandwidth) and phase-II (200 MHz bandwidth) have very different red noise parameters. On the basis of the current understanding of red noise parameters, we have devised search period ranges for two phases of GHRSS surveys: 0.5$-$100~s for phase-I and 0.1$-$100~s for phase-II. 

From the implementation of concurrent processing with FFA and FFT search in the GHRSS survey, we report the discovery of two pulsars, one of which having $<$ 1\% duty-cycle with a 1.1~s period, was only seen with the FFA search. Currently with a year of 
FFA processing, the number of long period pulsars (P $>$ 1 s) in the GHRSS survey increases to 5 from 2. Hopefully, similar search methodology in the major pulsar surveys can reveal the true population of long period pulsars and possibly populate the regions near death line in $P-\dot{P}$ plane, which can help to probe the conditions favouring the ceasing of the radio emission from pulsars.    


\section{Acknowledgement}
We acknowledge the support of the Department of Atomic Energy, Government of India,  under project no.12-R\&D-TFR-5.02-0700.  The GMRT is run by the National Centre for Radio Astrophysics of the Tata Institute of Fundamental Research, India. We acknowledge the support of GMRT telescope operators for the GHRSS survey observations.  We thank Dipanjan Mitra from NCRA-TIFR for helpful discussions on topics related to this paper. We also acknowledge the discussion with Naomi van Jaarsveld from South African Astronomical Observatory, while building the GHRSS FFA search pipeline using \texttt{RIPTIDE}.  MAM is supported by NSF Physics Frontiers Center award number 2020265 and by NSF AAG award number 2009425. Portions of this work performed at NRL were supported by the Office of Naval Research.

\bibliography{sample63}{}
\bibliographystyle{aasjournal}

\end{document}